\documentclass[a4paper,11pt]{article}
\usepackage{pdflscape}
\usepackage{amsmath,nccmath}
\usepackage{lipsum}
\usepackage{cite}
\usepackage{amssymb}
\usepackage{dcolumn}
\usepackage{bm}
\usepackage{xcolor}
\usepackage{epsfig}
\usepackage{amsfonts}
\usepackage{graphicx}
\usepackage{subfigure}
\usepackage{dcolumn}
\usepackage{indentfirst}
\usepackage{authblk}
\usepackage{slashed}
\usepackage{appendix}
\usepackage{makecell}
\usepackage{multirow}

\usepackage{booktabs}
\AtBeginDocument{
\heavyrulewidth=.08em
\lightrulewidth=.05em
\cmidrulewidth=.03em
\belowrulesep=.65ex
\belowbottomsep=0pt
\aboverulesep=.4ex
\abovetopsep=0pt
\cmidrulesep=\doublerulesep
\cmidrulekern=.5em
\defaultaddspace=.5em
}

\usepackage[colorlinks=false]{hyperref}
\hypersetup{citecolor=Blue}
\begin{document}

\vspace{80pt}
\centerline{\LARGE Holographic imaginary potential of a quark antiquark pair
\vspace{10pt}}
{\LARGE \quad\quad\quad\quad\quad in the presence of
 gluon condensation}

\vspace{40pt}
\centerline{
Sara Tahery,$^{a}$
\footnote{E-mail: s.tahery@impcas.ac.cn }
Xurong Chen, $^{b}$
\footnote{E-mail: xchen@impcas.ac.cn}
Zi-qiang Zhang$^{c}$
\footnote{E-mail: zhangzq@cug.edu.cn}}
\vspace{30pt}

{\centerline {$^{a,b}${\it Institute of Modern Physics, Chinese
Academy of Sciences, Lanzhou 730000,  China }} \vspace{4pt}
{\centerline {$^{b}${\it University of Chinese Academy of
Sciences, Beijing 100049, China }} \vspace{4pt} {\centerline
{$^{b}${\it  Guangdong Provincial Key Laboratory of Nuclear
Science, Institute of Quantum Matter,}}\vspace{4pt}
{\centerline {{\it  South China Normal
University, Guangzhou 510006, China}}
 \vspace{4pt} {\centerline
{$^{c}${\it School of Mathematics and Physics, China University of
Geosciences, Wuhan 430074, China }}

\vspace{40pt}

\begin{abstract}
For a moving heavy quark antiquark ($Q\bar{Q}$) in a quark gluon plasma (QGP), we use gauge/gravity duality to study both real and imaginary parts of the potential (Re$V_{Q\bar{Q}}$ and Im$V_{Q\bar{Q}}$ respectively) in a gluon condensate (GC) theory. The complex potential is derived from the Wilson loop by considering the thermal fluctuations of
the worldsheet of the Nambu-Goto holographic string. We calculate Re$V_{Q\bar{Q}}$ and Im$V_{Q\bar{Q}}$ in both cases where the axis of the moving $Q\bar{Q}$ pair is transverse and parallel with respect to its direction of movement in the plasma. Using the renormalization scheme for the Re$V_{Q\bar{Q}}$, we find that the inclusion of GC increases the dissociation length while rapidity has the opposite effect. While for the Im$V_{Q\bar{Q}}$, we observe that by considering the effect of GC, the Im$V_{Q\bar{Q}}$ is generated for larger distance thus decreasing quarkonium dissociation, while rapidity has opposite effect. In particular, as the value of GC decreases in the deconfined phase, the Im$V_{Q\bar{Q}}$ is generated for smaller distance thus enhancing quarkonium dissociation, and at high temperatures it is nearly not modified by GC, consistent with previous findings of the entropic force.
\end{abstract}

\newpage

\tableofcontents

\section{Introduction}\label{Introduction}
The heavy ion collisions at Relativistic Heavy Ion Collisions
(RHIC) and Large Hadron Collider (LHC) have produced a new state
of matter  called QGP
\cite{starcol2005,phenix2005,RHIC2005}. One of the most
experimental signatures of QGP formation is the dissociation of
quarkonia, like $c\bar{c}$ in the medium \cite{Shuryak1980,
mst,msd,mmk,gac}. Most studies over the past years have found that
the main mechanism responsible for this dissociation is color
screening \cite{mats1986}, however, recent studies suggest a more
important reason than the screening, the Im$V_{Q\bar{Q}}$ \cite{Thakur2017}. Moreover, this
quantity could be used to estimate the thermal width which is an
important subject in QGP \cite{nbma,ybm}. Calculations of the Im$V_{Q\bar{Q}}$
associated with QCD and heavy ion collisions were performed for
static pairs using pQCD \cite{mlop,Laine2007}. However,
theoretical analysis and experimental data demonstrate that QGP is
strongly coupled \cite{Shuryak2004}, then non-perturbative methods
are needed. The equation of state of the QGP at zero and finite
temperature are given in \cite{Borsanyi2014,Bazavov2014} and the
Im$V_{Q\bar{Q}}$ was studied in \cite{arth,gaca,gcs} by using non-perturbative
lattice QCD. Note that lattice QCD is very useful but  still
very difficult to use  to study real-time QCD dynamics. An
alternative method to study different aspects of QGP is the
AdS/CFT correspondence.

AdS/CFT conjecture originally relates the type IIB string theory
on $AdS_5 \times S^5$ space-time to the four-dimensional
$\mathcal{N}=4$ SYM gauge theory \cite{adscft}. In a holographic
description of AdS/CFT, a strongly coupled field theory at the
boundary of the AdS space is mapped onto the weakly coupled
gravitational theory in the bulk of AdS \cite{holo}. Although SYM
differs from QCD in many properties (e.g., at zero temperatures,
SYM is a conformal theory with no particle spectrum while QCD is a
confining theory with a sensible particle interpretation), it
reveals some qualitative features of QCD in strongly coupled
regime at non-zero temperatures (e.g., both theories describe hot,
non-Abelian plasmas with qualitatively similar hydrodynamic
behavior \cite{KT}). Given that, one could use the AdS/CFT
correspondence to study various aspects of QGP. An example of the
most known AdS/CFT calculations is the ratio of shear viscosity to entropy density \cite{KT,GD}.

On the other hand, the AdS/CFT correspondence has been generalized
to more realistic QCD, e.g, bottom-up holographic model
\cite{JE,LD}. The bottom-up approach begins with a
five-dimensional effective field theory somehow motivated by
string theory and tries to fit it to QCD as much as possible. In
the gravitational dual of QCD, the presence of probe branes in the
AdS bulk breaks the conformal symmetry and sets the energy scales,
so corrections in $AdS_5$ are useful to find more phenomenological
results. Holographic GC model
\cite{keh1999,Cski2007,Bak2004} is a type of bottom-up model with
phenomenological applicability as an effective model for the QGP.
It is known that GC is a measure of the non-perturbative physics
in zero-temperature QCD \cite{Shifman}, it is also an order
parameter for (de)confinement hence could be a condition for the
phase transition (the usual order parameter for the deconfinement
transition at finite temperature is the Polyakov loop. Also the
Wilson loop can be used to identify the (de)confined phases of
pure YM theory by its area law behavior). Although there is no
order parameter for the real-world QGP, GC model may be useful to
study the nonperturbative nature of the QGP \cite{lee89,MD2003,
Mil2007,QCDcon,action,Colangelo2013,Castorina2007,kim2007,Kopnin2011,chen2019},
such as in RHIC physics \cite{Brown2007}. In the mentioned
references it is shown that QCD sum rules is used to study the
nonperturbative physics of the strong interaction at zero
temperature. In this approach, the nonperturbative nature of the
vacuum is summarized in terms of quark and gluon condensates. To
study hot systems, one generalizes the technique to finite
temperature. The non-perturbative physics remaining even at high
temperatures, is manifested through the non-vanishing of some of
the vacuum condensates. Furthermore, using the GC model, the
thermodynamic properties of the system are discussed in
\cite{kim2007}, in which the well-known Stephan-Boltzmann law with
no condensation case can be recovered, also the energy density for
high temperature is given but for low temperature other back
ground is dominating. In the same reference, the dilaton (or GC)
contribution of the energy momentum tensor is identified as the
difference of the total and the thermal gluon, the GC contributes
negative energy which is a reminiscent of the zero temperature
result of Shifman, Vainstein and Zakharov \cite{Shifman}. In both
cases, the negativeness is coming from the renormalization. Also
in \cite{kim2007} the pressure, the trace anomaly and the entropy
density are given in presence of GC, as it is
expected, the entropy in condensed state is less than that in
thermal state. On the other hand, various observables or
quantities have been studied in Holographic GC model. For
instance, the GC dependency of the heavy quark
potential was studied in \cite{Kim2008} and the results indicate
that the potential becomes deeper as the GC in the
deconfined phase decreases and the mass of the quarkonium drops
near the deconfinement temperature $T_c$ (lattice QCD results
\cite{boyed96, boyedmiller} show that the GC appears
a drastic change near $T_c$). The GC dependency of
the jet quenching parameter and drag force was analyzed in
\cite{zhangdrag} and it was found that the two quantities both
decrease as the GC decreases in the deconfined
phase, indicating that the energy loss decreases near $T_c$. In
\cite{zhangPLB803} it is shown that the dropping GC
near $T_c$ increases the entropic force and thus enhances the
quarkonium dissociation.

The aim of this work is to analyze the Im$V_{Q\bar{Q}}$ of moving
$Q\bar{Q}$ in holographic GC model using the
world-sheet thermal fluctuations method \cite{Noronha2009,fn,sif}.
Note that the effect of the medium in the motion of a $Q\bar{Q}$
should be taken into account and the pair's rapidity through the
plasma has some effects on their interactions. In the LHC, the
heavy quarkonia are not only produced in large numbers but also
with high momenta so it is essential to consider the effect of
bound state speed on dissociation \cite{Ejaz2008}. As discussed in
\cite{sif}, one can adopt the saddle point approximation and
discuss the motion of a heavy quarkonia in a plasma and its
imaginary potential. The imaginary potential of the $Q\bar{Q}$ results from
the effect of thermal fluctuations due to the interactions between
quarks and the strongly coupled medium. By integrating out thermal
long wavelength fluctuations in the path integral of the
Nambu-Goto action in the background spacetime, a formula for the
imaginary part of the Wilson loop can be found in this approach
that is valid for any gauge theory dual to classical gravity.
Already, different holographic models were applied to study the
Im$V_{Q\bar{Q}}$ of the $Q\bar{Q}$
\cite{patra2015,saha2019,aliakbari,sara1412,ziqiang,sara2015,zhao2020,alba2008,Bitaghsir2013,Bitaghsir2015,
Bitaghsir2016,Braga2016,Erlich2005,
 Teramond2005,kim114008,Escobedo2014,Feng2020,Hayata2013,Burnier2009,Dudal2015,Braga2018, Bellantuono2019,Bellantuono2017}. It is worth to mention that
in \cite{sif} the Im$V_{Q\bar{Q}}$ of moving $Q\bar{Q}$ is considered in
strongly coupled plasma (associated with pure AdS), while in this
work we extend it to the AdS with GC. In \cite{patra2015} the Im$V_{Q\bar{Q}}$
of moving $Q\bar{Q}$ in OKS model (a type of holographic model) is
discussed, by delving into the solution beyond the critical
separation of the pair, which leads to the complex-valued string
configurations, not the world-sheet thermal fluctuations method as
we use here. In \cite{saha2019} the Im$V_{Q\bar{Q}}$ of moving $Q\bar{Q}$ in a
soft wall model with broken conformal invariance and finite
chemical potential is discussed. As holographic examples, such as
\cite{sif}, \cite{patra2015,saha2019,aliakbari} and
\cite{Bitaghsir2015} considered the moving cases, and they
concluded that increasing rapidity leads to decreasing the
dissociation length, implying the pair will be dissolved easier
into a moving medium compared to the static medium, consistent
with our findings here. However, from EFT point of view, the
reference \cite{Escobedo1304} explored the in-medium modifications
of heavy quarkonium states moving through a medium for two
plausible situations: $m_Q \gg 1/r \gg T \gg E \gg m_D$ and $m_Q
\gg T \gg 1/r,m_D \gg E$, results are relevant for moderate
temperatures and for studying dissociation, respectively. The
width decreases with the velocity for the former situation whereas
for the latter regime the width increases monotonically with the
velocity.

This paper is organized as follows: in the next section we study
the Re$V_{Q\bar{Q}}$  for both cases in which the dipole
moves transversely and parallel to the dipole axis. We proceed to
calculate the Im$V_{Q\bar{Q}}$ for the two cases mentioned in
the section \ref{se:ImV}. Section \ref{sec:Conclusions} contains
results and conclusions.

\section{Potential of moving  $Q\bar{Q}$ in presence of gluon condensation }\label{ReV}
In this section we evaluate the Re$V_{Q\bar{Q}}$ energy of the moving
$Q\bar{Q}$. The heavy quark potential (the vacuum interaction
energy) is related to the vacuum expectation value of the Wilson
loop \cite{wilson,Gervais,polygaugering} as,
\begin{equation}\label{W}
\lim_{_\mathcal{T}\longrightarrow 0}\langle W(\mathcal{C})\rangle_{0}\sim e^{i\mathcal{T} V_{Q\bar{Q}}(L)},
\end{equation}
where $\mathcal{C}$ is a rectangular loop of spatial length L and
extended over $\mathcal{T}$ in the time direction. The expectation
value of the Wilson loop can be evaluated in a thermal state of
the gauge theory with the temperature $T$. From this point of view
$V_{Q\bar{Q}}(L)$ is the heavy quark potential at finite
temperature and its imaginary part defines a thermal decay
width. To estimate the Im$V_{Q\bar{Q}}$  mentioned, one can use worldsheet fluctuations of the Nambu-Goto action \cite{Noronha2009}.\\
Note that although the Nambu-Goto action on top of a background
with a non-trivial dilaton field contains a coupling of this field
with the Ricci scalar as it is shown in \cite{Gursoy2007}, but the
contribution of this term is small. Therefore the well-known
modified holographic model introducing the GC in
the boundary theory is given by the following background action,
\begin{equation}\label{main action}
S=-\frac{1}{2k^2} \int d^5x \sqrt{g} \left(\mathcal{R}+ \frac{12}{L^2}-\frac{1}{2}\partial_{\lambda}\varphi \partial^{\lambda}\varphi\right),
\end{equation}
where $k$ is the gravitational coupling in $5$-dimensions,
$\mathcal{R}$ is Ricci scalar, $L$ is the radius of the asymptotic
$AdS_5$ spacetime, and  $\varphi$ is a massless scalar which is
coupled with the gluon operator on the boundary. By considering
the following ansatz the equations of the above action could be
solved, \cite{keh1999,Cski2007,Bak2004},
\begin{equation}\label{eq:metric}
ds^2= \frac{R^2}{z^2}( A(z)dx_i^2-B(z)dt^2+dz^2),
\end{equation}
where in this dilaton black hole background , $A(z),B(z), f$ are
defined as,
\begin{eqnarray}\label{eq:AB}
A(z)&=&(1+fz^4)^{\frac{f+a}{2f}}(1-fz^4)^{\frac{f-a}{2f}},\nonumber\\
B(z)&=&(1+fz^4)^{\frac{f-3a}{2f}}(1-fz^4)^{\frac{f+3a}{2f}},\nonumber\\
f^2&=&a^2+c^2,
\end{eqnarray}
$a$ is related to the temperature by $a=\frac{(\pi T)^4}{4}$ and the dilaton field is given by,
\begin{equation}\label{eq:dilaton}
\phi(z)=\frac{c}{f}\sqrt{\frac{3}{2}} \ln \frac{1+fz^4}{1-fz^4}+\phi_0.
\end{equation}

In (\ref{eq:metric}) $i = 1, 2, 3$ are orthogonal spatial boundary
coordinates, $z$ denotes the 5th dimension, radial coordinate and
$z = 0$ sets the boundary. $\phi_0$ in (\ref{eq:dilaton}) is a
constant. We work in the unit where $R=1$. Note that the dilaton
black hole solution is well defined only in the range
$0<z<f^{-1/4}$, where $f$ determines the position of the
singularity and $z_f$ behaves as an IR cutoff. For $a = 0$, it
reduces to the dilaton-wall solution. Meanwhile, for $c = 0$, it
becomes the Schwarzschild black hole solution. Also, for both
solutions, expanding the dilaton profile near $z = 0$ will give,
\begin{equation}\label{dilatonexpansion}
\phi(z)=\phi_0+\sqrt{6}\, c\, z^4+....
\end{equation}
$c$ is nothing but the holographic GC parameter.
As discussed in \cite{kim2007}, there exists a Hawking page
transition between the dilaton wall solution and dilaton blackhole
solution at some critical value of $a$. So the former is for the
confined phase, while the latter describes the deconfined phase.
The term $G_2$ is the vacuum expectation value of the
operator $\frac{\alpha_s}{\pi} G^a_{\mu\nu}G^{a,\mu\nu}$ where
$G^a_{\mu\nu}$ is the gluon field strength tensor. A non-zero
trace of the energy-momentum tensor appears in a full quantum
theory of QCD. The anomaly implies a non-zero GC
which can be calculated
 as \cite{Colangelo2013,Leutwyler1992,Castorina2007},
\begin{equation}\label{G}
\Delta G_2(T)= G_2(T)-G_2(0)=-(\varepsilon(T)-3\,P(T)),
\end{equation}
where $G_2(T)$ denotes the thermal GC,
$G_2(0)$, being equal to the condensate value at the deconfinement
transition temperature, is the zero temperature condensate value,
$\varepsilon(T)$ is the energy density, $P(T)$ is the pressure of the
QGP system.

To account for the effect of rapidity, one starts from a reference
frame where the plasma is at rest and the dipole is moving with a
constant velocity so it can be boosted to a reference frame where
the dipole is at rest but the plasma is moving past it \cite{sif}.
Consider a $Q\bar{Q}$ pair moving along $x_3$ direction with
rapidity $\eta$. Correspondingly, we can consider a reference
frame in which the plasma is at rest and the dipole moves with a
constant rapidity $-\eta$ in the $x_3$ direction. The usual way is
to boost the pair into a frame which is at rest while the hot wind
of QGP moves against it, so one can consider the following boost
to a reference frame in which the dipole is at rest but the plasma
is moving past it \cite{sif},
\begin{eqnarray}\label{eq:boost}
dt\rightarrow dt \cosh \eta - dx_3 \sinh \eta \nonumber \\
dx_d \rightarrow -dt \sinh \eta + dx_3 \cosh \eta,
\end{eqnarray}
if we transform the metric (\ref{eq:metric}) with (\ref{eq:boost}) we obtain,
\begin{eqnarray}\label{eq:metricboosted}
ds^2&=& \frac{1}{z^2}\Big{(} A(z)\, dx_{i}^2+[\cosh^2 \eta \, A(z)-\sinh^2 \eta \, B(z)] \, dx_3^2-[\cosh^2 \eta \, B(z)-\sinh^2 \eta \, A(z)]\, dt^2\nonumber \\
&-&2[A(z)- B(z)]\,\sinh \eta \cosh \eta \, dx_3 \, dt +dz^2\Big{)},
\end{eqnarray}
from now on, we can consider the dipole  in the gauge theory,
which has a gravitational dual with  metric
(\ref{eq:metricboosted}). In continue, we will consider the dipole
in two different ways, one transverse to the wind and one parallel
to it as,
\begin{equation}\label{directiontrans}
t=\tau, \quad x_1=\sigma, \quad  z=z(\sigma),\quad \text{transverse to the wind},
\end{equation}
\begin{equation}\label{directionpara}
t=\tau, \quad x_3=\sigma, \quad  z=z(\sigma),\quad \text{parallel to the wind},
\end{equation}
remind that in static gauge $z=z(\sigma,\tau)=z(\sigma).$

\subsection{Pair alignment transverse to the  wind, Re$V_{Q\bar{Q}}$ }
Consider the dipole is moving in the $x_1$ direction that is
transverse to the $x_3$. From (\ref{directiontrans}) the spacetime
target functions $X^\mu$ are $t=\tau$, $x_1=x=\sigma$,
$x_2=x_3=constant$,$ z(\sigma)$. The heavy $Q\bar{Q}$ potential
energy $V_{Q\bar{Q}}$ of this system is related to the expectation
value of a rectangular Wilson loop,
\begin{equation}\label{eq:wilsonloohol}
\langle W(\mathcal{C}) \rangle \sim e^{-i S_{str}},
\end{equation}
$S_{str}$ is the classical Nambu-Goto action of a string in the
bulk, so, on the dilaton black hole background, the Nambu-Goto
action is given by \cite{Kim2008},
\begin{equation}
\label{eq:nambugoto}
S_{str} = \frac{1}{2\pi \alpha'} \int d\sigma d\tau e^{\frac{\phi(z)}{2}}\sqrt{-det(G_{\mu\nu} \partial_{\alpha} X^{\mu} \partial_{\beta} X^{\nu}}).
\end{equation}
The dilaton factor $e^{\frac{\phi(z)}{2}}$ accounts for the fact
that $G_{\mu\nu}$ is the five-dimensional Einstein metric while
the target space metric is in the string frame. Therefore, the
non-zero coupled dilaton field (\ref{eq:dilaton}) to the
background metric (\ref{eq:metric}) should be considered when
writing the Nambu-Goto action of a test string.

Plugging back $S_{str}$ (\ref{eq:nambugoto}) in
(\ref{eq:wilsonloohol}) we extract the Re$V_{Q\bar{Q}}$ of $Q\bar{Q}$. Starting
from the metric (\ref{eq:metricboosted}), dilaton field (\ref{eq:dilaton}) , and the spacetime
target functions  we get,
\begin{equation}
\label{eq:nambugotoperpstatic} S_{str}=\frac{\mathcal{T}}{2\pi
\alpha'} \int_{-L/2}^{L/2} d\sigma \sqrt{f_1(z)\cosh^2 \eta-
f_2(z)\sinh^2 \eta  + (f_3(z)\cosh^2 \eta- f_4(z)\sinh^2
\eta)z'^2(\sigma) }.
\end{equation}
The quarks are located at $x_3 = \frac{L}{2}$ and $x_3 = -\frac{L}{2}$, $z'=\frac{dz}{d \sigma}$  and we defined,
\begin{eqnarray}\label{eq:f}
f_1(z)&=&\frac{\omega^2(z)}{z^4}\, A(z)\,B(z),\nonumber\\
f_2(z)&=&\frac{\omega^2(z)}{z^4} \, A^2(z),\nonumber\\
f_3(z)&=&\frac{\omega^2(z)}{z^4}\, B(z),\nonumber\\
f_4(z)&=&\frac{\omega^2(z)}{z^4}\, A(z),
\end{eqnarray}
and,
\begin{equation}\label{eq:lambda}
\omega(z)= e^{\frac{\phi(z)}{2}}= (\frac{1+fz^4}{1-fz^4})^{\frac{c}{f}\sqrt{\frac{3}{8}}}.
\end{equation}
 We also write,
\begin{eqnarray}\label{F,G}
 F(z)&=&f_1(z)\cosh^2 \eta- f_2(z)\sinh^2 \eta, \nonumber\\
 G(z)&=& f_3(z)\cosh^2 \eta- f_4(z)\sinh^2 \eta.
\end{eqnarray}
So action (\ref{eq:nambugotoperpstatic}) could be written as,
\begin{equation}\label{eq:finalaction}
S_{str} = \frac{\mathcal{T}}{2\pi \alpha'} \int_{-L/2}^{L/2} d\sigma \sqrt{F(z) + G(z)z'^2(\sigma) }.
\end{equation}
The action depends only on $\sigma=x$ and the associated
Hamiltonian is a constant of the motion. With the corresponding
position of the deepest position in the bulk being $z_*$,
Hamiltonian is,
\begin{equation}\label{hamil}
H=\frac{F(z)}{\sqrt{F(z) + G(z)z'^2(\sigma) }}=cte=\sqrt{F(z_*)}.
\end{equation}
From the Hamiltonian (\ref{hamil}), we can write the equation of
motion for $z(x)$ as,
\begin{equation}
\label{eq:eomperp} \frac{dz}{dx}=\Big{[}\frac{F(z)}{G(z)}\Big{(}
\frac{F(z)}{F(z_*)}-1  \Big{)}  \Big{]}^{\frac{1}{2}}.
\end{equation}
Therefore,
\begin{equation}
\label{eq:x,L}
 dx=\Big{[}\frac{F(z)}{G(z)}\Big{(} \frac{F(z)}{F(z_*)}-1  \Big{)} \Big{]}^{-\frac{1}{2}} dz,
\end{equation}
and we can relate $L$ to $z_*$ as follows,
\begin{equation}
\label{eq:x,z}
 \frac{L}{2}=\int_{0}^{z_*}\Big{[}\frac{F(z)}{G(z)}\Big{(} \frac{F(z)}{F(z_*)}-1  \Big{)} \Big{]}^{-\frac{1}{2}} dz.
\end{equation}
From (\ref{eq:x,z}) we  find the length of
the line connecting both quarks as,
\begin{eqnarray}\label{eq:L,z}
 L&=&2\sqrt{F(z_*)}\int_{0}^{z_*}\Big{[} \frac{G(z)}{F(z)(F(z)-F(z_*))}\Big{]}^{\frac{1}{2}} dz.
 \end{eqnarray}
In the literature \cite{LiuPRL,LiuJHEP} the maximum value of the
above length has been used to define a dissociation length for the
moving $Q\bar{Q}$ pair, where the dominant configuration for
$S_{str}$ is two straight strings (two heavy quarks) running from
the boundary to the horizon.

If we put (\ref{eq:eomperp}) in (\ref{eq:finalaction}) the action is written as follows,
\begin{equation}
\label{eq:Sperpnreg0} S_{str} = \frac{\mathcal{T}}{\pi \alpha'}
\int_{0}^{z_*} dz \, \sqrt{G(z)} \sqrt{\frac{F(z)}{F(z_*)}}
\left[\frac{F(z)}{F(z_*)}-1 \right]^{-1/2}.
\end{equation}

Note that (\ref{eq:Sperpnreg0}) is divergent, which is
characteristic of Wilson loops, due to the fact that the string
must stretch from the bulk to the boundary in the holographic
approach. To regularize the above integral, one can subtract the
divergence in $S_{str}$ and obtain the regularized Wilson loop as
\cite{fn,sif}
\begin{equation}
\label{eq:Sperpreg} S^{reg}_{str} = \frac{\mathcal{T}}{\pi
\alpha'} \int_{0}^{z_*} dz \, \sqrt{G(z)}
\sqrt{\frac{F(z)}{F(z_*)}}  \left[\frac{F(z)}{F(z_*)}-1
\right]^{-1/2} - \frac{\mathcal{T}}{\pi \alpha'} \int_{0}^{\infty}
dz \sqrt{f^0_3(z)},
\end{equation}
where $f^0_3(z)=f_3(z)\mid_{a\longrightarrow 0}$ (quark self energy).
Finally, we proceed from  $\mathrm{Re}\,V_{Q\bar{Q}} = S^{reg}_{str}/\mathcal{T}$  to,
\begin{equation} \label{eq:Revz}
 \mathrm{Re}\,V_{Q\bar{Q}}  =
\frac{\sqrt{\lambda}}{\pi} \int_{0}^{z_*} dz \, \sqrt{G(z)}
\sqrt{\frac{F(z)}{F(z_*)}}  \left[\frac{F(z)}{F(z_*)}-1
\right]^{-1/2} - \frac{\sqrt{\lambda}}{\pi} \int_{0}^{\infty} dz
\sqrt{f^0_3(z)},
\end{equation}
\begin{figure}[h!]
\begin{minipage}[c]{1\textwidth}
\tiny{(a)}\includegraphics[width=8cm,height=6cm,clip]{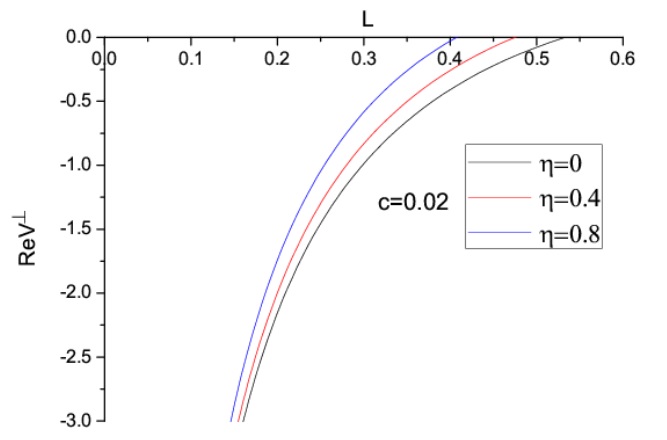}
\tiny{(b)}\includegraphics[width=8cm,height=6cm,clip]{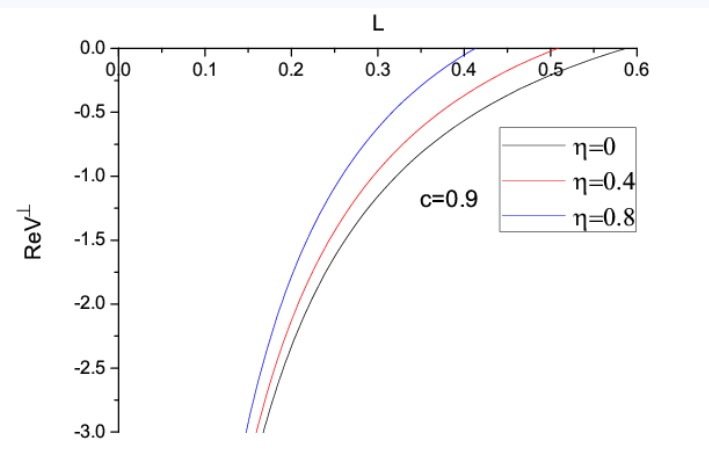}
\end{minipage}
\caption{$\mathrm{Re}\,V_{Q\bar{Q}}$ as a function of $L$ for a
$Q\bar{Q}$ pair oriented transverse to the  axis of the quarks,
from top to bottom for $\eta=0.8,\, 0.4,\,0$ respectively,
$T=200\,MeV$, in the presence of GC, for a)
$c=0.02$\,
 Ge$V^4$\, and\, b) $c=0.9$\, Ge$V^4$. }
\label{Revtrans}
\end{figure}
where $\lambda=\frac{1}{\alpha'^2}$ is the 't Hooft coupling of
the gauge theory. Figure \ref{Revtrans} shows the Re$V_{Q\bar{Q}}$ as a
function of $L$ with the $Q\bar{Q}$ pair oriented transverse to
the axis of the quarks, in the presence of GC. In this figure and
all other plots from now on, we consider $T=200\, MeV$, because at
low temperature the heavy quarkonia are hard to dissociate and as
the temperature increases the dissociation is more likely to
happen \cite{Kim2008}. The results show that increasing rapidity
leads to a decrease in dissociation length while $c$ has the
opposite effect.

\subsection{Pair alignment parallel to the wind, Re$V_{Q\bar{Q}}$}
In this step we consider that the dipole moves parallel to the
$x_3$ direction. From (\ref{directionpara}) the spacetime target
functions $X^{\mu}$ are $t=\tau, x_1=x_2=constant, x_3=x=\sigma, z(\sigma)
$. Using steps similar to (\ref{eq:finalaction}) we get the
action with the new worldsheet as,
\begin{equation}\label{eq:finalactionpar}
S_{str} = \frac{\mathcal{T}}{2\pi \alpha'} \int_{-L/2}^{L/2} d\sigma \sqrt{f_1(z) + G(z)z'^2(\sigma) },
\end{equation}
where $G(z)$ and $f_1(z)$ are defined as (\ref{F,G}) and (\ref{eq:f}).
Similar to the transverse case, we find the line connecting both quarks as,
\begin{eqnarray}\label{eq:L,zpar}
 L&=&2\sqrt{f_1(z_*)}\int_{0}^{z_*}\Big{[} \frac{G(z)}{f_1(z)(f_1(z)-f_1(z_*))}\Big{]}^{\frac{1}{2}} dz,
\end{eqnarray}
and the Re$V_{Q\bar{Q}}$  as,
\begin{equation}
\label{eq:Revzpar} \mathrm{Re}\,V_{Q\bar{Q}}  =
\frac{\sqrt{\lambda}}{\pi} \int_{0}^{z_*} dz \, \sqrt{G(z)}
\sqrt{\frac{f_1(z)}{f_1(z_*)}}  \left[\frac{f_1(z)}{f_1(z_*)}-1
\right]^{-1/2} - \frac{\sqrt{\lambda}}{\pi} \int_{0}^{\infty} dz
\sqrt{f^0_3(z)}.
\end{equation}
\begin{figure}[h!]
\begin{minipage}[c]{1\textwidth}
\tiny{(a)}\includegraphics[width=8cm,height=6cm,clip]{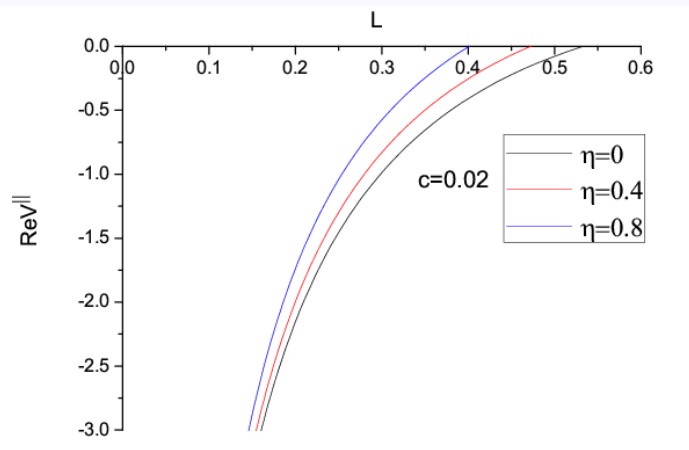}
\tiny{(b)}\includegraphics[width=8cm,height=6cm,clip]{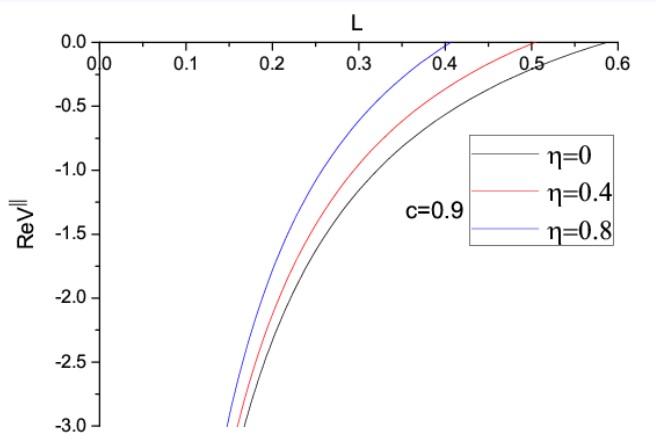}
\end{minipage}
\caption{$\mathrm{Re}\,V_{Q\bar{Q}}$
 as a function of $L$ for a $Q\bar{Q}$ pair oriented
parallel to the  axis of the quarks,  from top to bottom for
$\eta=0.8,\, 0.4,\,0$ respectively, $T=200\,MeV$, in the presence
of GC, for a) $c=0.02$\, Ge$V^4$\, and\, b)
$c=0.9$\, Ge$V^4$.} \label{Revpara}
\end{figure}

Figure \ref{Revpara} shows the Re$V_{Q\bar{Q}}$ as a function of $LT$ for some
choices of $\eta$ where $Q\bar{Q}$ pair oriented parallel to the
axis of the quarks, in presence of GC. Similar to previous case,
increasing rapidity leads to decreasing the dissociation length
while $c$ has the opposite effect.
\begin{figure}[h!]
\begin{center}$
\begin{array}{cccc}
\includegraphics[width=10cm]{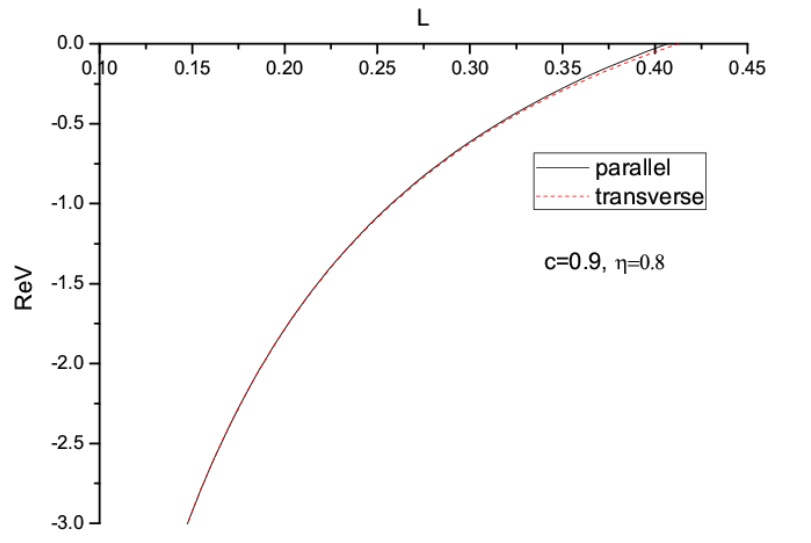}
\end{array}$
\end{center}
\caption{$\mathrm{Re}\,V_{Q\bar{Q}}$
 as a function of $L$, for fixed value of $\eta$ and fixed value of $c$, $T=200\,MeV$, as a comparison between the parallel and the transverse cases.
 The solid black line shows parallel case and the dashed red line shows transverse case. }
\label{fig:VpvsVs}
\end{figure}

Figure \ref{fig:VpvsVs} shows a comparison between the Re$V_{Q\bar{Q}}$  for  the parallel and the transverse cases.
Although the difference is not significant, the plots show that
the effect of the GC is slightly stronger for the
parallel case. In other words, increasing $c$ increases the
dissociation length in both the transverse and the parallel cases
(previous figures), this effect appears stronger when the dipole
moves parallel to the axis of the quarks.

\section{Imaginary potential of moving $Q\bar{Q}$ in presence of gluon condensation }\label{se:ImV}
In this section, we calculate the Im$V_{Q\bar{Q}}$ using the
thermal worldsheet fluctuations method for both the transverse and
parallel cases.
\subsection{ Pair alignment transverse to the  wind, Im$V_{Q\bar{Q}}$ }
Consider the effect of worldsheet fluctuations around the classical configuration $r=\frac{1}{z}$,
\begin{equation}\label{zfluc}
r(x) = r_*(x) \rightarrow r(x) = r_*(x) + \delta r (x),
\end{equation}
then the fluctuations in the partition function should be considered  as follows,
\begin{equation}\label{Zwithfluc}
Z_{str} \sim \int \mathcal{D} \delta r(x) e^{i S_{NG} (r_*(x) + \delta r (x))}.
\end{equation}
Hence there is an imaginary part of the potential in the action. Dividing the interval  of $x$ into $2N$ points (where $N\longrightarrow\infty$) we obtain,
\begin{equation}\label{Zwithflucfin}
Z_{str} \sim \lim_{N\to \infty}\int d [\delta r(x_{-N})] \ldots d[ \delta r(x_{N})]  \exp{\left[ i \frac{\mathcal{T} \Delta x}{2 \pi \alpha'} \sum_j \sqrt{\tilde{G}
\, r_j^{'2} +\tilde{F}}\right]},
\end{equation}
where $\tilde{G}$ and $\tilde{F}$ are functions of $r_j$. We
expand $r_*(x_j)$ around $x=0$ and keep only terms up to second
order of it because thermal fluctuations are important around
$r_\ast$ which means $x=0$,
\begin{equation}\label{zstarexpan}
r_*(x_j) \approx r_\ast + \frac{x_j^2}{2} r_*''(0),
\end{equation}
 considering small fluctuations  we have,
\begin{equation}\label{Fwithfluc}
\tilde{F} \approx \tilde{F}_* + \delta r \tilde{F}'_* + r_*''(0) \tilde{F}'_* \frac{x_j^2}{2} + \frac{\delta r^2}{2} \tilde{F}''_*,
\end{equation}
where $\tilde{F}_\ast\equiv \tilde{F}(r_\ast)$  and $\tilde{F}'_\ast\equiv \tilde{F}'(r_\ast)$. The action is written as,
\begin{equation}\label{actionwithroots}
S^{NG}_j = \frac{\mathcal{T} \Delta x}{2 \pi \alpha'} \sqrt{C_1 x_j^2 + C_2},
\end{equation}
where $C_1$ and $C_2$ are given as follows,
\begin{equation}\label{C1}
C_1 = \frac{r_*''(0)}{2} \left[ 2 \tilde{G}_* r_*''(0) + \tilde{F}_*' \right],
\end{equation}
\begin{equation}\label{C2}
C_2 = \tilde{F}_* + \delta r \tilde{F}'_* + \frac{\delta r^2}{2} \tilde{F}''_*,
\end{equation}
 to have $ Im V_{Q\bar{Q}}\neq 0$, the function in the square root (\ref{actionwithroots})  should be negative. Then, we consider the j-th contribution to $Z_{str}$ as,
\begin{equation}\label{Ij}
I_j \equiv \int\limits_{\delta r_{j min}}^{\delta r_{j max}} D(\delta r_j) \, \exp{\left[ i \frac{\mathcal{T} \Delta x}{2 \pi \alpha'} \sqrt{C_1 x_j^2 + C_2} \right]},
\end{equation}
\begin{equation}\label{Ddeltaz}
D(\delta r_j) \equiv C_1 x_j^2 + C_2(\delta r_j),
\end{equation}
 \begin{equation} \label{deltaz}
\delta r = - \frac{\tilde{F}'_*}{\tilde{F}''_*},
\end{equation}
so, $ D(\delta r_j)<0 \Longrightarrow  -x_*<x_j<x_*$ leads  to an imaginary part in the square root. We write,
\begin{equation}\label{xstar}
x_* = \sqrt{\frac{1}{C_1}\left[\frac{\tilde{F}'^2_*}{2\tilde{F}''_*} - \tilde{F}_* \right]},
\end{equation}
if the square root is not real we should take $ x_*=0$. With all these conditions we can approximate $D(\delta r) $ by $ D(-\frac{\tilde{F}'_{\ast}}{\tilde{F}"_{\ast}})$ in $ I_j$ as,
\begin{equation}\label{Ijapprox}
I_j \sim \exp \left[ i \frac{\mathcal{T} \Delta x}{2 \pi \alpha'} \sqrt{C_1 x_j^2 + \tilde{F}_* - \frac{\tilde{F}'^2_*}{2\tilde{F}''_*}} \right].
\end{equation}
The total contribution to the imaginary part, will be available with a continuum limit. So,
\begin{equation}\label{ImV}
\mathrm{Im} \, V_{Q\bar{Q}} = -\frac{1}{2\pi \alpha'} \int\limits_{|x|<x_*} dx \sqrt{-x^2 C_1 - \tilde{F}_* + \frac{\tilde{F}'^2_*}{2\tilde{F}''_*}},\,
\end{equation}
which leads to,
\begin{equation}\label{ImVr}
\mathrm{Im} \, V_{Q\bar{Q}} = -\frac{1}{2 \sqrt{2} \alpha'} \sqrt{\tilde{G}_*} \left[\frac{\tilde{F}'_*}{2\tilde{F}''_*}-\frac{\tilde{F}_*}{\tilde{F}'_*} \right].
\end{equation}
Note that (\ref{ImVr}) is the Im$V_{Q\bar{Q}}$ with the $r$
coordinate. Changing the variable back to the coordinate
$z=\frac{1}{r}$ according to our background, we will
have,
\begin{equation}\label{ImVz}
\mathrm{Im} \, V_{Q\bar{Q}} =
-\frac{1}{2 \sqrt{2} \alpha'} \sqrt{G_*}
z^2_*\left[\frac{F_*}{z^2_*F'_*}-\frac{z^2_* F'_*}{4z^3_* F'_*+2
z_*^4\, F''_*} \right],
\end{equation}
 where $F$ is again a function of $z$. In (\ref{ImVz}) the following condition should be satisfied for the square root,
 \begin{equation}\label{ImVcondition}
 \frac{B(z_*)}{A(z_*)}> tanh^2 \eta.
 \end{equation}
\begin{figure}[h!]
\begin{minipage}[c]{1\textwidth}
\tiny{(a)}\includegraphics[width=8cm,height=6cm,clip]{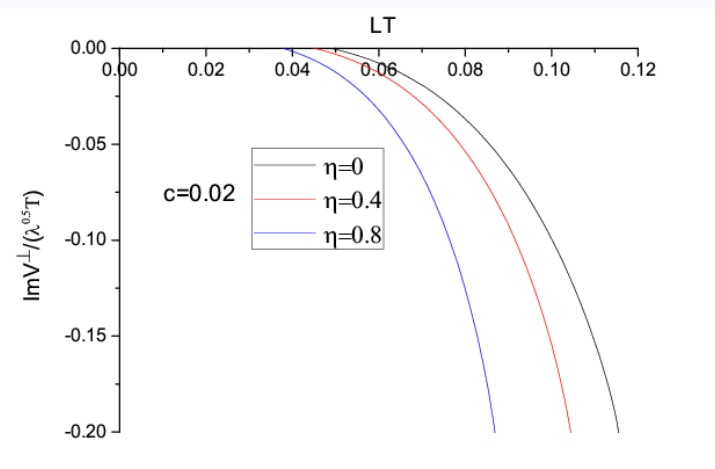}
\tiny{(b)}\includegraphics[width=8cm,height=6cm,clip]{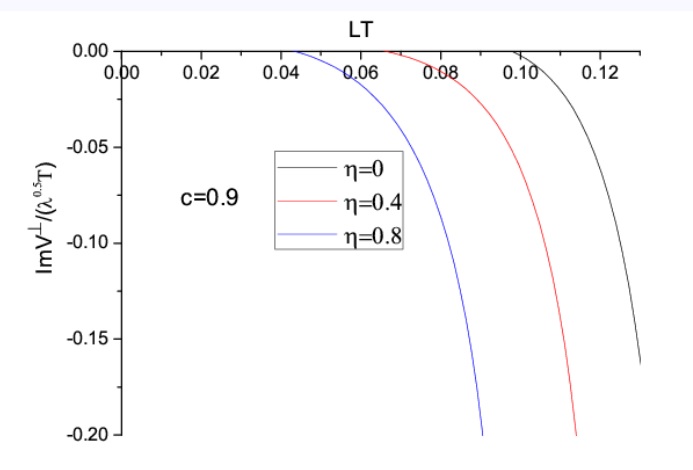}
\end{minipage}
\caption{Im$V_{Q\bar{Q}}$ as a function of $LT$ for a $Q\bar{Q}$
pair oriented transverse to the  axis of the quarks, from left to
right for $\eta=0.8,\, 0.4,\, 0$ respectively, $T=200\,MeV$, in
the presence of GC,
 for a) $c=0.02$\, Ge$V^4$\, and\, b) $c=0.9$\, Ge$V^4$.} \label{Imvtrans}
\end{figure}

Figure \ref{Imvtrans} shows the Im$V_{Q\bar{Q}}$ as a function
of $LT$ for some choices of $\eta$ where $Q\bar{Q}$ pair oriented
transverse to the axis of the quarks, in the presence of GC. With increasing rapidity the Im$V_{Q\bar{Q}}$ is generated for smaller 
values of $LT$, implying quarkonium melts more easily, consistent
with the results of \cite{Fadafan2016}. Also, by comparing the two panels, one finds $c$ has
opposite effects, i.e. by increasing GC, the Im$V_{Q\bar{Q}}$ is generated for larger distance thus decreasing quarkonium dissociation. 

\subsection{Pair alignment parallel to the wind, Im$V_{Q\bar{Q}}$}
Taking action (\ref{eq:finalactionpar}) and using the same
approach we followed to find (\ref{ImVz}), we get the Im$V_{Q\bar{Q}}$ of a pair moving  parallel to the axis of the quarks as,
\begin{equation}\label{ImVzpar}
\mathrm{Im} \, V_{Q\bar{Q}} = -\frac{1}{2 \sqrt{2} \alpha'} \sqrt{G_*} z^2_*\left[\frac{f_{1*}}{z^2_*f'_{1*}}-\frac{z^2_* f'_{1*}}{4z^3_* f'_{1*}+2 z_*^4\, f''_{1*}} \right].
\end{equation}
\begin{figure}[h!]
\begin{minipage}[c]{1\textwidth}
\tiny{(a)}\includegraphics[width=8cm,height=6cm,clip]{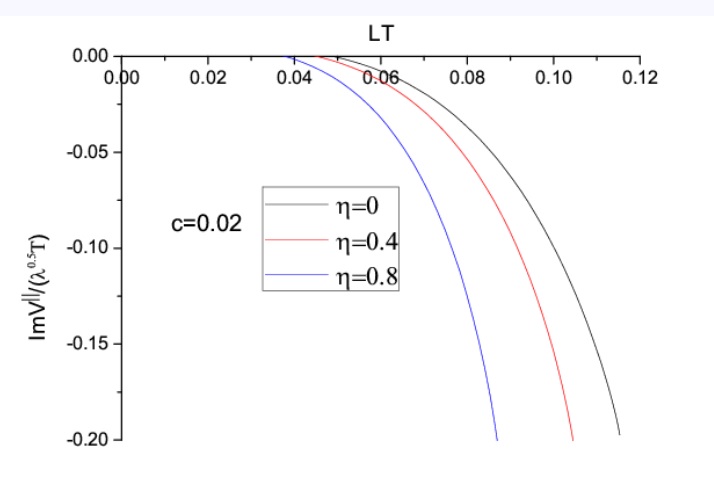}
\tiny{(b)}\includegraphics[width=8cm,height=6cm,clip]{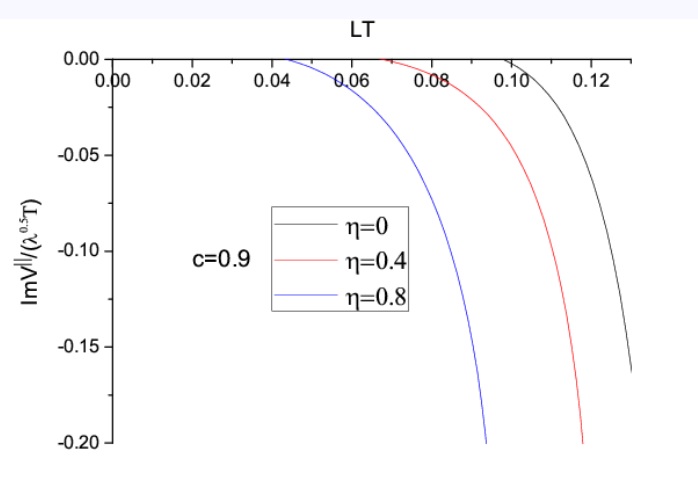}
\end{minipage}
\caption{Im$V_{Q\bar{Q}}$
 as a function of $LT$ for a $Q\bar{Q}$ pair oriented
parallel to the  axis of the quarks, from left to right for $\eta=0.8,\, 0.4,\, 0$ respectively, $T=200\,MeV$, in the presence of GC, for a)  $c=0.02$\, Ge$V^4$\,
and\, b) $c=0.9$\, Ge$V^4$.}
\label{Imvpara}
\end{figure}

Figure \ref{Imvpara} shows the Im$V_{Q\bar{Q}}$ as a function
of $LT$ for some choices of $\eta$ where $Q\bar{Q}$ pair oriented
parallel to the axis of the quarks, in the presence of GC. The $c$ parameter in the gravitational dual,
introduces the GC in the QCD side of duality.
Therefore with increasing GC the Im$V_{Q\bar{Q}}$ starts to become non-zero for larger values of $LT$. Similar to the transverse case, these effects are the opposite of
the rapidity effects.
 \begin{figure}[h!]
\begin{center}$
\begin{array}{cccc}
\includegraphics[width=10cm]{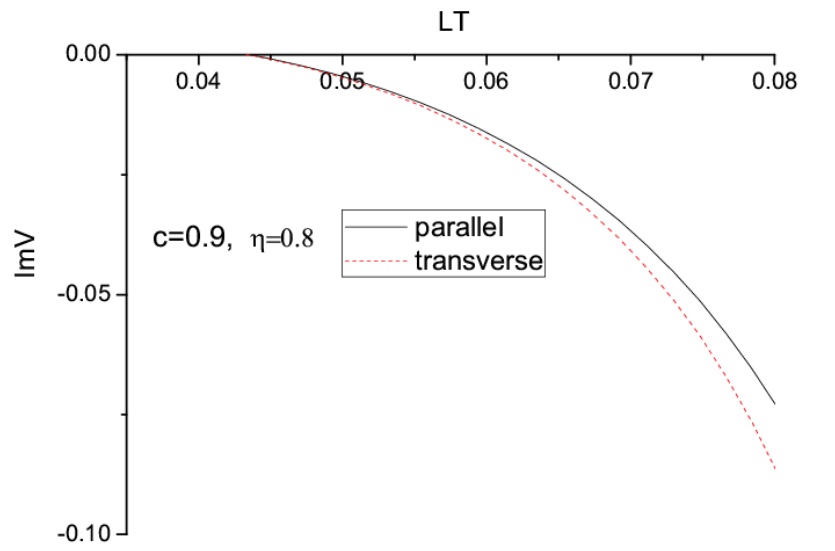}
\end{array}$
\end{center}
\caption{Im$V_{Q\bar{Q}}$ as a function of $LT$, for fixed value
of $\eta$ and fixed value of $c$, $T=200\,MeV$, as  a comparison
between the parallel and the transverse cases. The (right) black
line
 shows parallel case and the (left)  red line shows transverse case. }
\label{fig:ImVpvsImVs}
\end{figure}
Figure \ref{fig:ImVpvsImVs} shows a comparison between the
Im$V_{Q\bar{Q}}$ for the parallel and the
transverse cases. Similar to Re$V_{Q\bar{Q}}$ in figure \ref{fig:VpvsVs}, the
plots show that the effect of the GC is stronger
for the parallel case. While the magnetic field
\cite{ziqiang} and the chemical potential effects were more
important for the transverse case, in the parallel case the GC has a stronger impact.

\section{Conclusions}\label{sec:Conclusions}
In this work, we investigated the heavy quark potential of a moving $Q\bar{Q}$ pair in a plasma considering the effect of GC.
We calculated the Re$V_{Q\bar{Q}}$ and Im$V_{Q\bar{Q}}$ for the cases where the axis of the moving pair is transverse and
parallel with respect to its rapidity in the plasma, respectively. For the Re$V_{Q\bar{Q}}$, we used the renormalization scheme proposed in \cite{fn,sif}
and observed the inclusion of GC decreases Re$V_{Q\bar{Q}}$ and
increases the dissociation length, opposite to the effect of the rapidity. For the Im$V_{Q\bar{Q}}$, we adopted the
world-sheet thermal fluctuations method \cite{Noronha2009,fn,sif} and found increasing GC, the Im$V_{Q\bar{Q}}$ is generated for larger distance thus decreasing quarkonium dissociation, while rapidity has opposite effect, consistent with the findings of the entropic force \cite{zhangPLB803}.
However, we have to admit that the model considered here has a shortcoming: GC is constant implying temperature dependence is absent.
An analogous situation happens in \cite{Kim2008}. But the existing research, e.g., \cite{boyed96, boyedmiller} indicates that the GC appears
a drastic change near $T_c$. In addition, \cite{Colangelo2013} shows that, when T (temperature) is not very high, GC strongly depends on T and $\mu$ (chemical
potential), and at high temperatures, GC becomes independent of T and $\mu$. From the above analysis, one may infer that as GC decreases in the deconfined phase, Im$V_{Q\bar{Q}}$ is generated for smaller distance thus enhancing quarkonium dissociation, and at high temperatures Im$V_{Q\bar{Q}}$ is nearly not modified by GC. It should be noted that we could not give a concrete conclusion on the
Im$V_{Q\bar{Q}}$ at intermediate or low temperatures. To solve this problem we would need
to study the competitive effects of GC, $\mu$ and T  on the Im$V_{Q\bar{Q}}$ and also the correlation among those three. We hope to work on this topic in future.\\
\\
\textbf{Acknowledgement}\\
Authors would like to thank Kazem Bitaghsir Fadafan for useful
comments. This work was supported by Strategic Priority Research
Program of Chinese Academy of Sciences (XDB34030301). Sara Tahery
is suppoted by the CAS President International Fellowship
Initiative, PIFI (2021PM0065) and Younger Scientist Scholarship
(QN2021043004, funding number E11l631KR0).

\end{document}